\documentclass[showpacs,preprintnumbers,amsmath,amssymb,floatfix,prd]{revtex4}
\usepackage{epsf,epsfig}

\usepackage{graphicx}% Include figure files
\usepackage{bm}% bold math
\usepackage{amsfonts}
\usepackage{amsmath}
\usepackage[utf8]{inputenc}

\usepackage{amssymb}
\usepackage{graphicx}
\usepackage{epsfig}
\usepackage{amsfonts}
\usepackage[dvipsnames, usenames]{xcolor}

\begin{document}

\title{
Spinning and excited black holes in 
Einstein-scalar-Gauss-Bonnet theory
}

\author{{Lucas G. Collodel$^{1,2}$}}
 \email{lucas.gardai-collodel@uni-tuebingen.de}
\author{{Burkhard Kleihaus$^2$}}
\author{{Jutta Kunz$^2$}}
\author{{Emanuele Berti$^3$}}
\affiliation{
$^1$ Theoretical Astrophysics, Eberhard Karls University of T\"ubingen, 
D-72076 T\"ubingen, Germany\\
$^2$Institut f\"ur Physik, Universit\"at Oldenburg, Postfach 2503,
D-26111 Oldenburg, Germany\\
$^3$ Department of Physics and Astronomy, Johns Hopkins University, 
Baltimore, MD 21218 USA
}

\date{\today}

 \begin{abstract} 
    We construct rotating black holes in Einstein-scalar-Gauss-Bonnet theory
with a quadratic coupling function.
We map the domain of existence of the rotating fundamental solutions,
we construct radially excited rotating black holes (including their
existence lines), and we show that there are angularly excited
rotating black holes. The bifurcation points of the radially and
angularly excited solutions branching out of the Schwarzschild solution
follow a regular pattern.
  \end{abstract}

\maketitle

\section{Introduction}

In General Relativity (GR) the existence of asymptotically flat
black holes is subject to severe constraints, often termed
no-hair theorems 
(see e.g.~\cite{Chrusciel:2012jk,Herdeiro:2015waa,Cardoso:2016ryw}).
In generalized theories of gravity, on the other hand,
less restrictions may arise and thus these may lead 
to interesting new kinds of asymptotically flat black holes,
that carry hair
(see e.g.~\cite{Berti:2015itd}).
These hairy black holes might in fact represent contenders to explain
current astrophysical observations
\cite{Berti:2018cxi,Berti:2018vdi}.

A particularly interesting class of generalized theories of gravity
is represented by metric theories with higher curvature terms.
Such theories arise, for instance, in the low-energy limit
of string theory, where these higher curvature terms
are accompanied by a scalar field, the dilaton
\cite{Gross:1986mw,Metsaev:1987zx}.
The resulting black holes then carry scalar hair,
as shown for the case of a Gauss-Bonnet (GB) term
coupled to a dilaton \cite{Kanti:1995vq,Torii:1996yi,Guo:2008hf,Pani:2009wy}.

The physical properties of nonrotating hairy black holes can differ
significantly from their GR counterparts, the Schwarzschild black holes.
In particular, the presence of the scalar field
will give rise to additional branches in the black hole quasinormal mode spectrum
\cite{Pani:2009wy,Ayzenberg:2013wua,Blazquez-Salcedo:2016enn,Blazquez-Salcedo:2017txk}.
When set into rotation, the quadrupole moments of these hairy black holes can
exhibit large deviations 
from those of Kerr black holes,
and their angular momentum may even exceed the Kerr bound, $j=J/M^2=1$
\cite{Pani:2011gy,Ayzenberg:2014aka,Maselli:2015tta,Kleihaus:2011tg,Kleihaus:2014lba,Kleihaus:2015aje}.
In constrast, the shadows of hairy black holes 
and their X-ray reflection spectrum
will be very close to those of Kerr black holes
\cite{Cunha:2016wzk,Zhang:2017unx}. 

Einstein-dilaton-Gauss-Bonnet gravity is characterized by an exponential coupling function
$f(\phi)$ to the GB term, whose exponent is linear in the
dilaton field $\phi$. Therefore Schwarzschild or Kerr black holes
are not solutions of the theory: they are only approached
asymptotically.
If, however, one allows for other choices of the coupling function
(the simplest being a quadratic coupling function $f(\phi)\propto \phi^2$)
Schwarzschild and Kerr black holes can be solutions of the theory,
and an interesting new phenomenon can arise: curvature-induced
spontaneous scalarization of black holes
\cite{Doneva:2017bvd,Silva:2017uqg,Antoniou:2017acq,Antoniou:2017hxj,
Blazquez-Salcedo:2018jnn,Doneva:2018rou,Silva:2018qhn,Cunha:2019dwb,Macedo:2019sem}.

Spontaneous scalarization was first observed in neutron stars
within scalar-tensor theory. Here the instability arises when
the product $-\beta_0 T$, where $\beta_0$ is the effective linear
matter-scalar coupling and $T$ is the trace of the stress-energy tensor,
is larger than some critical value 
\cite{Damour:1993hw}:
spontaneous scalarization in neutron stars is induced by couplings
with matter (see also \cite{Mendes:2016fby,Motahar:2017blm}).
Later it was realized that spontaneous scalarization can occur
for charged black holes in Einstein-Maxwell-scalar (EMs) theory,
for certain choices of the scalar coupling function 
and coupling strength
\cite{Stefanov:2007eq,Doneva:2010ke,Herdeiro:2018wub,Myung:2018vug,Myung:2018jvi,Myung:2019oua}.
This ``charge-induced'' spontaneous scalarization presents many similarities with the case of curvature-induced spontaneous 
scalarization of black holes
\cite{Doneva:2017bvd,Silva:2017uqg,Antoniou:2017acq,Antoniou:2017hxj,
Blazquez-Salcedo:2018jnn,Doneva:2018rou,Silva:2018qhn,Cunha:2019dwb,Macedo:2019sem}.

In Einstein-scalar-Gauss-Bonnet (EsGB) gravity, the presence of black holes
with scalar hair that is spontaneously induced by curvature is associated
with instabilities of Schwarzschild black holes.
In particular, as the coupling constant $\lambda/M$ is varied,
a set of bifurcation points arises, where branches of
scalarized black holes emerge
\cite{Doneva:2017bvd,Silva:2017uqg}.
Labelling these bifurcation points by the integer $n$,
the scalarized solutions on the $n$th branch possess $n$ radial nodes.
Thus, besides the fundamental ($n=0$) scalarized black hole,
one can have radially excited modes with $n>0$ . With every new bifurcation point
the Schwarzschild black hole gains another unstable mode
\cite{Blazquez-Salcedo:2018jnn}. The stability of the
fundamental static solution depends on the coupling function 
\cite{Doneva:2017bvd,Silva:2017uqg,Antoniou:2017acq,Antoniou:2017hxj,
Blazquez-Salcedo:2018jnn,Doneva:2018rou,Silva:2018qhn,Cunha:2019dwb}
and on self-interaction terms, if they are present \cite{Macedo:2019sem}.

Recently, Ref.~\cite{Cunha:2019dwb} studied the fundamental ($n=0$) solution
for rotating BHs in EsGB theory with a 
``Gaussian'' coupling function of the form $e^{-\phi^2}$,
as well as its domain of existence and various of its physical properties.
The domain of existence 
is quite broad for small rotation rates (as expected from the static
solution), but it becomes narrower as rotation increases.
This fact has been exploited in calculations of the shadow of such
black holes, which might be used to put a bound on the coupling constant
 \cite{Cunha:2019dwb}.

Here we consider the static and rotating black holes of EsGB theory
with a simple quadratic coupling function. We explore the domain of 
existence of the fundamental rotating black holes and consider
their first radial excitations. Moreover, we show that the scalarized
static and rotating black holes
also possess angular excitations (labelled by an angular
integer $l$).
We determine the bifurcation points  of the lowest excitations
and determine the existence lines of some of the resulting
radially and angularly excited rotating black holes.
In Section~\ref{sec:framework} we describe the theory
and the general properties of axially symmetric EsGB black holes.
In Section~\ref{sec:numerics} we present our numerical results,
and in Section~\ref{sec:conclusions} we outline possible directions
for future work.

\section{General framework}
\label{sec:framework}

\subsection{Action  }
 
The action of EsGB gravity is
\begin{eqnarray}  
\label{act}
S=\frac{1}{16 \pi}\int d^4x \sqrt{-g} \left[R - \frac{1}{2}
 (\partial_\mu \phi)^2
 + f(\phi) R^2_{\rm GB}   \right],
\end{eqnarray} 
where 
$\phi$ is a (real) scalar field, $f(\phi)$ is the coupling function
of the theory, and
\begin{eqnarray} 
\label{GB}
 R^2_{\rm GB} = R_{\mu\nu\rho\sigma} R^{\mu\nu\rho\sigma}
- 4 R_{\mu\nu} R^{\mu\nu} + R^2
\end{eqnarray} 
is the Gauss-Bonnet invariant, which 
would not yield any modifications of the Einstein equations when
$f(\phi)$ is a constant,
because it corresponds to a boundary term in the action.
This is no longer the case if the GB invariant couples to dynamical matter fields.
Note that here and below we use geometrical units ($c=G=1$).

Here we consider the coupling function 
\cite{Silva:2017uqg,Blazquez-Salcedo:2018jnn}
\begin{equation}
f(\phi) = \frac{\lambda^2}{8} \phi^2 ,
%\alpha \phi^2 
\label{quad_cf}
\end{equation}
i.e., a purely quadratic coupling. We will compare our results with those for a Gaussian coupling
\begin{equation}
f(\phi) = \frac{\lambda^2}{12} \left(1 - e^{-3 \phi^2/2} \right) ,
%\alpha \phi^2 
\label{exp_cf}
\end{equation}
which was studied in \cite{Doneva:2017bvd,Blazquez-Salcedo:2018jnn,Cunha:2019dwb}.

Varying the action~(\ref{act}) with respect to 
the metric $g_{\mu\nu}$, we obtain
the  generalized Einstein equations 
with contributions from the GB term
\begin{eqnarray}
\label{EGB-eq}
&& E_{\mu\nu}=G_{\mu\nu}
-\frac{1}{2} T_{\mu\nu}^{(\phi)}
%-\frac12\biggl[\nabla_{\mu}\phi\nabla_{\nu} \phi -\frac12 g_{\mu\nu}(\nabla\phi)^2\biggr]
%\nonumber \\
%&& \hspace{10mm}
+f(\phi)\Bigl[H_{\mu\nu}
+4(\gamma^2\nabla^{\rho}\phi\nabla^{\sigma}\phi
-\gamma\nabla^{\rho}\nabla^{\sigma}\phi)P_{\mu\rho\nu\sigma}\Bigr]
=0,
\end{eqnarray}
where
\begin{eqnarray}
\nonumber
&&G_{\mu\nu}= R_{\mu\nu}-\frac{1}{2}g_{\mu\nu}R,~~~
%{\rm with}~~
T_{\mu\nu}^{(\phi)}=\nabla_{\mu}\phi\nabla_{\nu} \phi -\frac12 g_{\mu\nu}(\nabla\phi)^2,~~
\\
&&H_{\mu\nu}= 2\bigl(RR_{\mu\nu}-2R_{\mu \rho}R^{\rho}_{~\nu}
-2R^{\rho\sigma}R_{\mu\rho\nu\sigma}
+R_{\mu}^{~\rho\sigma\lambda}R_{\nu\rho\sigma\lambda}\bigr)
-\frac{1}{2}g_{\mu\nu}R^2_{\rm GB},
\\
\nonumber
&& P_{\mu\nu\rho\sigma}=
R_{\mu\nu\rho\sigma}+2g_{\mu[\sigma}R_{\rho]\nu}
+2g_{\nu[\rho}R_{\sigma]\mu} +Rg_{\mu[\rho}g_{\sigma]\nu}.
\label{EGB:eq}
\end{eqnarray}
In the above relations, 
we denote by $P_{\mu\nu\rho\sigma}$ 
the divergence free part of the Riemann tensor,
$i.e.$
$\nabla_\mu P^{\mu}_{~\nu\rho\sigma}=0$.
Obviously, the equations~(\ref{EGB-eq}) can be written 
in an Einstein-like form 
\begin{eqnarray}
\label{EGB-eq1}
 G_{\mu\nu}=\frac{1}{2} T_{\mu\nu}^{\rm (eff)}~,
 \end{eqnarray} 
where we have introduced an effective energy-momentum tensor 
that has acquired a contribution arising from the GB term
%\eb{The acronym GBd was never introduced - introduce, or use ``GB''}
\begin{eqnarray}
\label{Teff}
T_{\mu\nu}^{\rm (eff)}=T_{\mu\nu}^{(\phi)}
-2 T_{\mu\nu}^{\rm (GBd)},
\end{eqnarray} 
with
 \begin{eqnarray}
\label{TeffGB}
 T_{\mu\nu}^{\rm (GBd)}= H_{\mu\nu}
+4\nabla^{\rho}\nabla^{\sigma}f(\phi) P_{\mu\rho\nu\sigma} .
 \end{eqnarray}   
Variation of Eq.~(\ref{act}) with respect to the scalar field  leads
to a generalized Klein-Gordon equation, 
\begin{eqnarray}
\label{dil-eq}
\nabla^2 \phi +\frac{df}{d\phi}  R^2_{\rm GB}
 =0.
\end{eqnarray}

\subsection{The ansatz and equations of motion}

We would like to focus on stationary,
axially symmetric spacetimes possessing
two commuting Killing vector fields, $\xi$ and $\eta$, with
\begin{eqnarray}
\xi = \partial_t~~~ {\rm and}~~~\eta=\partial_\varphi
\end{eqnarray}
in a system of adapted coordinates.
Such spacetimes are typically described by a 
Lewis-Papapetrou--type ansatz  \cite{Wald:1984rg},
which satisfies the circularity condition and contains four unknown functions.
Here we employ the version of this ansatz originally introduced in
\cite{Kleihaus:2000kg},
with the parametrization 
\begin{eqnarray}
\label{metric}
ds^2=- b e^{F_0} dt^2 + e^{F_1} \left( d r^2+ r^2d\theta^2 \right) 
           +  e^{F_2} r^2\sin^2\theta (d\varphi-\frac{\omega}{r} dt)^2 ,
\end{eqnarray}
where $r$, $\theta$, and $\varphi$ are
``quasi-isotropic" spherical coordinates, and $t$ is the time coordinate. 
Here $b=(1-\frac{r}{r_{\rm H}})^2$ is an auxiliary function, 
and $r_{\rm H}$ denotes the isotropic horizon radius.
The metric functions $F_0$, $F_1$, $F_2$ and $\omega$ depend on 
the coordinates $r$ and $\theta$.
The scalar field is also a function of $r$ and $\theta$ only:
\begin{eqnarray}
\label{scalar}
\phi=\phi(r,\theta).
\end{eqnarray}

\subsection{Boundary conditions and asymptotic behavior}

{\bf~~ Large-$r$ asymptotics.}
We here consider solutions that approach a Minkowski spacetime background 
as $r\to \infty$. This implies the following boundary conditions:
\begin{eqnarray}
F_0(\infty)=F_1(\infty)=F_2(\infty)=\omega(\infty)=\phi(\infty)= 0
\ . \label{bc1a} 
\end{eqnarray}
Since the scalar field is massless, 
one can construct an approximate solution of the field equations 
(\ref{EGB-eq}) and (\ref{dil-eq}),
that is compatible with these asymptotic conditions 
as a power series in $1/r$.
%The leading order terms of such an expansion are:  
%\begin{eqnarray}
%\nonumber
%&&h  h e^{F_0}=1-\frac{2 M}{r} +\frac{2 M^2}{r^2}
%+\left(\frac{1}{3}M(C_1-4M^2)-2M_2 P_2(\cos\theta) \right)\frac{1}{r^3}
%+ O \left(\frac{1}{r^4}\right),
%\\
%\label{inf-expr}
%&&
%e^{F_1+F_0}=1 + \frac{C_1}{r^2} - \frac{M^2+2 C_1+D^2/4}{r^2}\sin^2{\theta}
%+ O\left( \frac{1}{r^3}\right),
%\\
%\nonumber
%&&e^{F_2+F_0}=1+\frac{C_1}{r^2} + O\left(\frac{1}{r^3}\right),
~~~
%\omega=-\frac{2 a M}{r^2}+\frac{6 a M^2}{r^3} + O\left(\frac{1}{r^4}\right),
%~~\phi=-\frac{D}{r} +O\left(\frac{1}{r^3}\right),
% \end{eqnarray}
%where $M$, $C_1$, $a$, $M_1$ and $D$ are free parameters, 
%while $P_2(\cos \theta)$ is a Legendre polynomial of 2nd degree.
%(note that $D$ can be interpreted as the dilaton charge).
%

\bigskip
 
%%%%%%%%%%%%%%%%%%%%%%%%%%%%%%%%%%%%%%%%%%%%%%%%%%%%%%%%%%%%%%%%%
%%%%%%%%%%%%%%%%%%%%%%%%%%%%%%%%%%%%%%%%%%%%%%%%%%%%%%%%%%%%%%%%%
%\subsubsection{Expansion on the event horizon}
%%%%%%%%%%%%%%%%%%%%%%%%%%%%%%%%%%%%%%%%%%%%%%%%%%%%%%%%%%%%%%%%%%
{\bf~~ Expansion on the event horizon.}
The event horizon of the (non-extremal) stationary black hole solutions
resides at a surface of constant radial coordinate, $r=r_{\rm H}>0$.
At a regular horizon the metric functions must satisfy
\begin{eqnarray}
 \label{bc-horizon} 
%\partial_r F_0(r_{\rm H})=\partial_r F_1(r_{\rm H})=\partial_r F_1(r_{\rm H})=0
\partial_r F_0(r_{\rm H})=\frac{1}{r_{\rm H}}\ , \ \ \
\partial_r F_1(r_{\rm H})=-\frac{2}{r_{\rm H}}\ , \ \ \
\partial_r F_2(r_{\rm H})=-\frac{2}{r_{\rm H}} \ , \ \ \ 
\omega(r_{\rm H})=\omega_{\rm H},
\end{eqnarray}
where $\omega_{\rm H}$ is a constant, 
while the condition imposed on the scalar field is
\begin{eqnarray}
\partial_r \phi(r_{\rm H})=0.
\end{eqnarray}
Again, it is possible to construct an approximate (power series) solution,
this time in terms of
\begin{eqnarray}
\delta=\frac{r}{r_{\rm H}}-1.
\end{eqnarray}
%For non-extremal solutions (the case considered 
%explicitly in this work),
%the first terms in the near horizon expansion read
%\begin{eqnarray}
%&&
%\label{horizon-expansion}
%f(r,\theta)=\delta^2 f_2(\theta) (1 -\delta) + O(\delta^4) ,
%%
%~~
%m(r,\theta)=\delta^2 m_2(\theta) (1 -3\delta) + O(\delta^4),
%\\
%&&
%%\label{phi_hor_red}
%\nonumber
%l(r,\theta)=\delta^2 l_2(\theta) (1 -3\delta) + O(\delta^4),
~~
%\omega(r,\theta)=\omega_{\rm H} (1 + \delta) + O(\delta^2), 
~~
%\phi(r,\theta)=\phi_{0}(\theta)  + O(\delta^2) ,
%\end{eqnarray}
%with $f_2$, $m_2$, $l_2$ and $\phi_0$ unspecified
%functions\footnote{
Note that, similar to the case of pure Einstein gravity,
the equation $E_{r}^\theta=0$ implies that the ratio $F_0/F_1$ is constant. 
This implies the constancy of the Hawking temperature, 
and also represents a supplementary condition which is used as further 
test of numerical accuracy. 

\bigskip

{\bf~~Behavior on the symmetry axis.}
Axial symmetry and regularity impose
the following boundary conditions for the metric functions
on the symmetry axis
(i.e. at $\theta=0,\pi$):
\begin{eqnarray}
\label{bc-axis}
& &\partial_\theta F_0|_{\theta={0,\pi}} =
   \partial_\theta F_1|_{\theta={0,\pi}} =
   \partial_\theta F_2|_{\theta={0,\pi}} =
   \partial_\theta \omega|_{\theta={0,\pi}} = 0 \ , 
\end{eqnarray}
while for the scalar field we impose
\begin{eqnarray}
\partial_\theta \phi|_{\theta={0,\pi}} =0. 
\end{eqnarray}
Near the symmetry axis it is possible to construct an approximate form 
of the solutions as a power series,
now in terms of $\theta$ (and $\pi-\theta$, respectively).
%For example, the first terms in such an expansion as $\theta\to 0$ reads
Further, the absence of conical singularities implies 
%\begin{eqnarray}
$F_1|_{\theta=0,\pi} =F_2|_{\theta=0,\pi}$.
%\end{eqnarray}
%\begin{eqnarray}
%&&
%\nonumber
%f(r,\theta)=\bar f_0(r)+\theta^2 \bar f_2(r) + O(\theta^4) ,
%%
%~
%m(r,\theta)=\bar m_0(r)+\theta^2 \bar m_2(r) + O(\theta^4) ,
%~
%l(r,\theta)=\bar l_0(r)+\theta^2 \bar l_2(r) + O(\theta^4) ,
%\\
%&&
%%
%\label{z-expansion}
%\omega(r,\theta)=\bar \omega_0(r)+\theta^2 \bar \omega_2(r) + O(\theta^4) ,
%~~
%\phi(r,\theta)=\bar \phi_0(r)+\theta^2 \bar \phi_2(r) + O(\theta^4) .
%\end{eqnarray}

All fundamental solutions discussed here are symmetric with respect to
reflection across the equatorial plane, $\theta=\pi/2$.
Therefore, in the numerical calculations, 
it is sufficient to consider the range
$0\leq \theta \leq \pi/2$ for the angular variable $\theta$.
Then the metric functions and the scalar field are required 
to satisfy Neumann boundary conditions in the equatorial plane:
\begin{eqnarray}
\label{bc-equator}
\partial_\theta F_0|_{\theta=\pi/2} =
 \partial_\theta F_1|_{\theta=\pi/2} =
  \partial_\theta F_2|_{\theta=\pi/2} =
   \partial_\theta \omega|_{\theta=\pi/2} =
	\partial_\theta \phi|_{\theta=\pi/2} =0.
\end{eqnarray}
Note that the scalar field of the first angular excitation
possesses odd parity. In this case the  boundary condition
in the equatorial plane reads $\phi|_{\theta=\pi/2} =0$.

\subsection{Physical properties}

Let us now briefly address the physical properties of these black holes.
Starting with the horizon properties, we note that
%the event horizon has spherical topology,
the metric of a spatial cross section of the horizon is
\begin{eqnarray}
\label{horizon-metric}
d\Sigma^2=h_{ij} dx^i dx^j=r_{\rm H}^2
\left.\left(e^{F_1} d\theta^2+e^{F_2} \sin^2\theta d\varphi^2\right)\right|_{r_{\rm H}}.
\end{eqnarray}
The Killing vector field
\begin{equation}
\chi = \partial_t - \frac{\omega_{\rm H}}{r_{\rm H}}\partial_{\varphi}
\  \label{chi} 
\end{equation}
is orthogonal to (and null on) the horizon \cite{Wald:1984rg}.
%Also, 
The boundary parameter $\omega_{\rm H} $ 
%which enters the event horizon boundary conditions  
defined in Eq.~(\ref{bc-horizon})
determines the horizon angular velocity $\Omega_{\rm H}$ 
\begin{eqnarray}
\label{OmegaH}
\Omega_{\rm H}=-\frac{\xi^2}{\xi\cdot \eta}=-\frac{g_{\varphi t}}{g_{tt}}\bigg|_{r_{\rm H}}=\frac{\omega_{\rm H} }{r_{\rm H}}  .
\end{eqnarray}

The Hawking temperature $T_{\rm H}={\kappa}/({2\pi})$
is obtained from the surface gravity $\kappa$ \cite{Wald:1984rg},
where $\kappa^2=-\frac{1}{2}(\nabla_a \chi_b)(\nabla^a \chi^b)|_{r_{\rm H}}$,
yielding
 \begin{eqnarray}
\label{TH}
T_{\rm H}=\frac{1}{2 \pi r_{\rm H}}e^{(F_0-F_1)/2}  \ .
 \end{eqnarray}
The horizon area of the black holes is given by
 \begin{eqnarray}
\label{AH}
A_{\rm H}=2\pi r_{\rm H}^2\int_0^\pi d\theta 
\sin\theta e^{(F_0+F_2)/2}.
 \end{eqnarray}
Black holes in GR possess an entropy 
which is a quarter of the horizon area \cite{Wald:1984rg}.
However, because of the scalar coupling to the GB term,
the entropy of the EsGB black holes acquires an extra contribution.
Following Wald \cite{Wald:1993nt},
the total entropy can then be given as an integral over the horizon
\begin{eqnarray}
\label{S-Noether} 
S=\frac{1}{4}\int_{\Sigma_{\rm H}} d^{2}x 
\sqrt{h}(1+ 2 f(\phi) \tilde R),
\end{eqnarray} 
where $h$ is the determinant of the induced metric on the horizon 
(defined in (\ref{horizon-metric})),
and $\tilde R$ represents the horizon curvature.
%Its explicit form for the Ansatz in this work reads
%%
%\begin{eqnarray}
%&&S=S_{\rm E}+S_{\rm GBd},~~{\rm with}~~S_{\rm E}= \frac{1}{2}\pi r_{\rm H}^2\int_0^\pi d\theta \frac{\sqrt{l_2 m_2}}{f_2},
%~~~{\rm and}~~
%%\end{eqnarray} 
%%\begin{eqnarray}
%\\
%\nonumber
%&&~~~S_{\rm GBd}= \frac{1}{2}\pi \alpha \int_0^\pi d\theta
%\frac{ e^{-\gamma \phi_0}}{ l_2^{3/2}m_2^{5/2}} 
%\bigg(
% 2m_2^2l_2'^2 \sin\theta-m_2l_2(-3\sin\theta l_2' m_2'+4m_2(2\cos\theta l_2'+\sin\theta l_2''))
%\\
%\nonumber
%&&{~~~~~~~~~~~~~~~~~}+l_2^2(8m_2^2\sin\theta-3\sin \theta m_2'^2+2m_2(3\cos \theta m_2'+\sin\theta m_2''))
%\bigg)
%,
%\end{eqnarray}
%with a prime denoting a derivative with respect to $\theta$.

Similar to GR solutions,
the total mass $M$ and the angular momentum $J$ are read from 
the asymptotic %sub-leading 
behavior of the metric functions
\begin{eqnarray}
\label{asym}
g_{tt} &=-e^{F_0}+e^{F_2}\omega^2 \sin^2 \theta
  =-1+\frac{2M}{r}+\dots,\\
g_{\varphi t}&=- e^{F_2}\omega^2 \sin^2 \theta=-\frac{2J}{r}\sin^2\theta+\nonumber \dots.  
\end{eqnarray}  
In addition, the solutions possess a scalar ``charge'' $D$, which is 
determined by the $1/r$ term of the far-field asymptotics of the scalar field.

\subsection{Numerical approach}

Let us briefly address the numerical approach employed
to construct the EsGB black holes.
The unknown metric and scalar field functions $(F_0, F_1, F_2, \omega; \phi)$ 
are obtained as solutions of a rather lengthy coupled set of
partial differential equations (PDEs),
subject to the associated set of boundary conditions,
guaranteeing regularity and asymptotic flatness.
%In particular, the set of PDEs
%involves first and second derivatives of the unknown functions.

The only non-trivial components of the generalized Einstein equations (\ref{EGB-eq}) 
are $E_t^t, E_r^r,E_\theta^\theta,E_\varphi^\varphi,E_\varphi^t$ and $E_r^\theta$.
Following \cite{Kleihaus:2000kg},
we divide the resulting six equations into two groups.
Four equations for the metric functions
are obtained from a suitable linear combination of
$E_t^t,E_\varphi^\varphi,E_\varphi^t$ and $E_r^r+E_\theta^\theta$.
The remaining two equations ($E_r^r-E_\theta^\theta$ and $E_r^\theta$)
represent constraints. 
We solve the four equations together with 
the scalar field equation (\ref{dil-eq}),
each containing more than 340 independent terms,
while we follow the two constraints
to check the accuracy of the numerical solutions.

%which diagonalizes
%\footnote{Note that, however, due to the contribution of 
%the Gauss-Bonnet term, 
%the final equations are not diagonal with respect to second derivatives
%of ${\cal F}_i$.} 
%the corresponding Einstein tensor parts
%with respect to $\hat O F_0$, $\hat O F_1 $,  $\hat O  F_2$ and  
%$\hat O {\omega}$, respectively (with the 
%operator $\hat O=\partial_{rr}+\frac{1}{r^2}\partial_{\theta\theta}$).
 
The domain of integration corresponds to the region outside the horizon.
We therefore introduce a new radial variable $x=1-r_{\rm H}/r$,
which maps the semi-infinite interval $[r_{\rm H},\infty)$ 
to the closed interval $[0,1]$.
%This leads to the following substitutions in the differential equations
%%\begin{eqnarray}
%$r  {\cal F}_{,r} \longrightarrow \frac{1}{r_{\rm H}} (1-x) {\cal F}_{,x}$
%and 
%$ r^2 {\cal F}_{,rr}   \longrightarrow \frac{1}{r_{\rm H}^2} 
%\big( (1-x)^2  {\cal F}_{,xx} - 2 (1-x) {\cal F}_{,x} \big) $
%%\end{eqnarray}
%for each of the functions.
%
We then discretize the equations on a non-equidistant grid in
$x$ and $\theta$. 
Typical grids used have sizes $91 \times 51$,
and cover the integration region
$0\leq x \leq 1$ and $0\leq \theta \leq \pi/2$.  
We perform the numerical calculations  
using the professional package FIDISOL/CADSOL 
\cite{Schoenauer:1989},
which is based on a Newton-Raphson method.   
%
%This code provides also an error estimate for each unknown function.
The typical  numerical error 
for the functions is estimated to be lower than $10^{-3}$. 

For each solution we provide three input parameters, 
$\lambda$, $r_{\rm H}$ and $\Omega_{\rm H}=\frac{\omega_{\rm H}}{r_{\rm H}}$.
After convergence has been reached,
the physical properties are computed from the numerical solutions.
In particular, the mass $M$ and the angular momentum $J$
are obtained from the asymptotic behavior of the solutions  (\ref{asym}),
whereas the horizon area $A_{\rm H}$, the entropy $S$ and the
temperature $T_{\rm H}$ are extracted from the horizon metric.

%\subsection{Known limits of the model}
 
%Before discussing the properties of the EGBd
%spinning BHs, it is useful
%to briefly review the properties of the solutions in two important limits
%of the general model.

\section{Results}
\label{sec:numerics}

\begin{figure}[t]
        \centering
\mbox{
\includegraphics[width=0.50\textwidth]{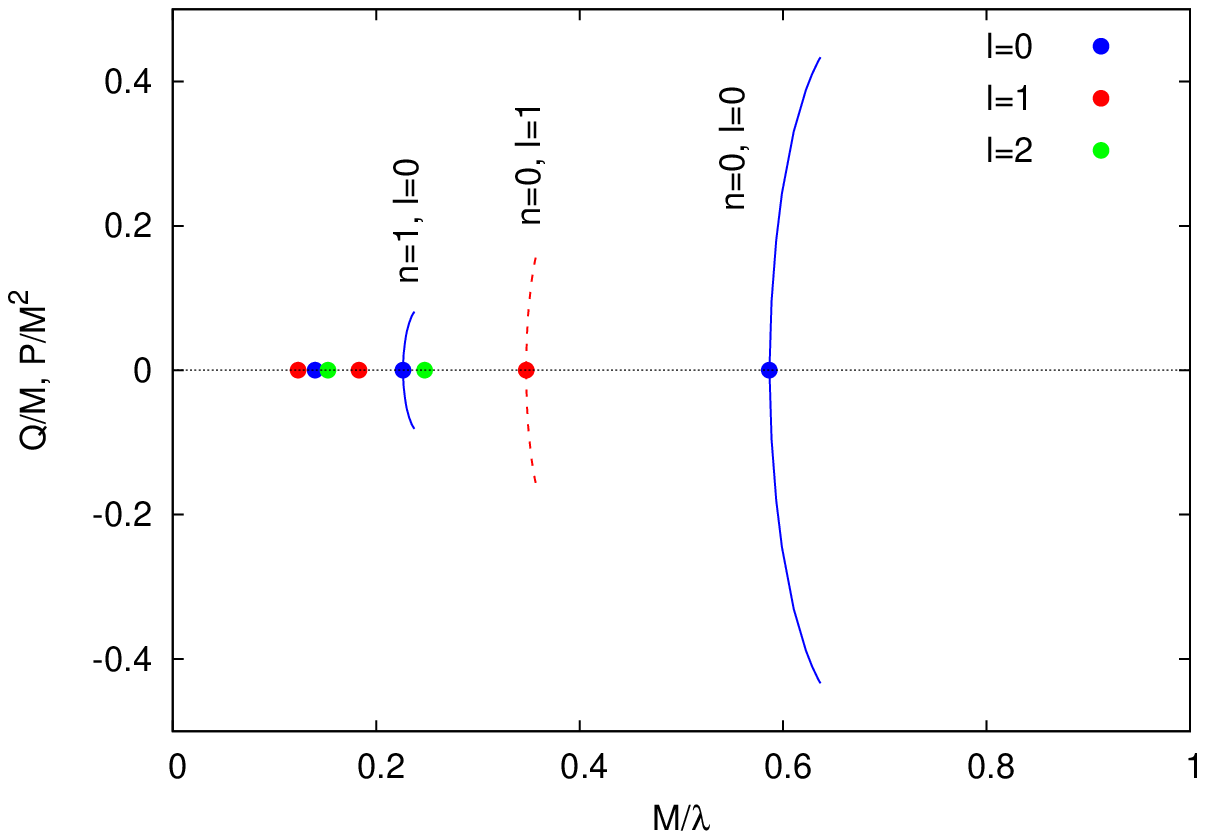}
\includegraphics[width=0.50\textwidth]{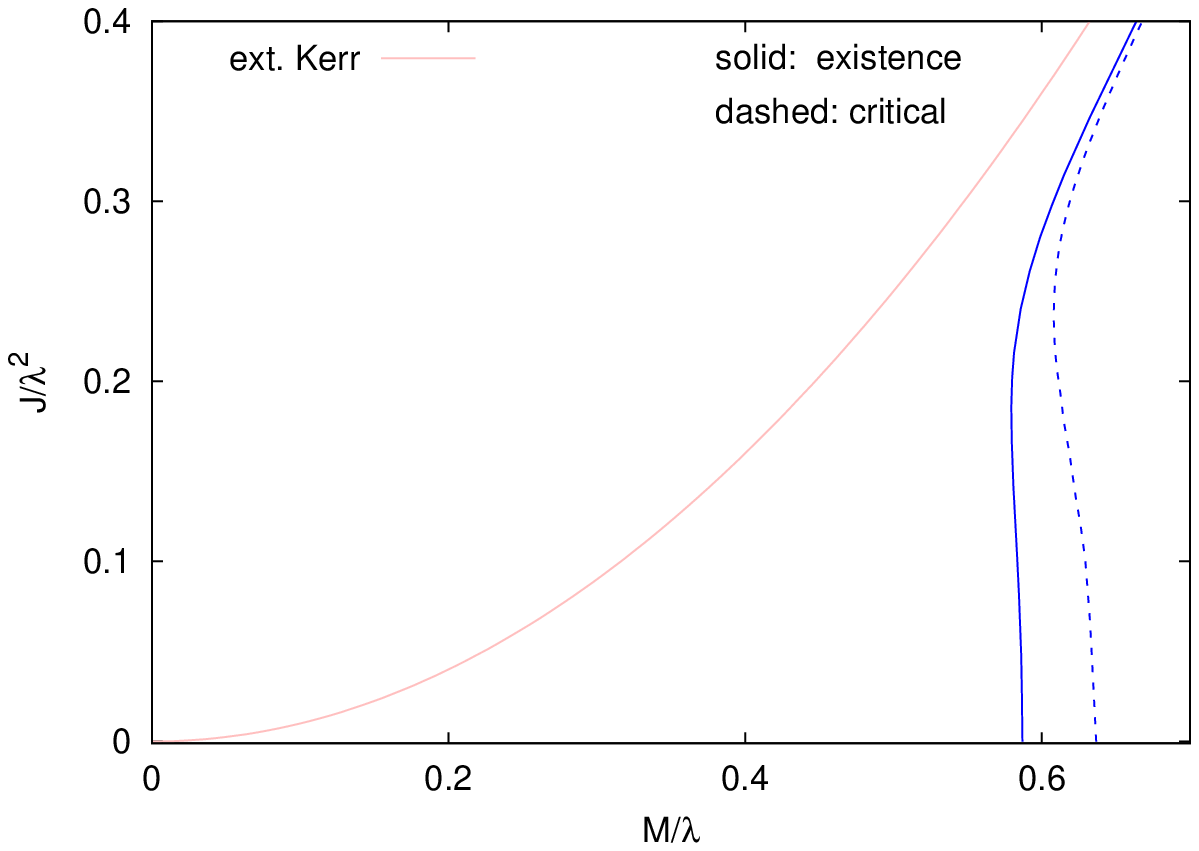}
}
        \caption{Domain of existence of scalarized black holes:
(a) fundamental and radially and angularly excited static solutions: 
%scaled scalar charge $Q/M$ vs scaled coupling $\lambda/M$
scaled scalar charge $Q/M$ vs scaled mass $M/\lambda$
for even $l$ ($l=0$: solid blue, $l=2$: green),
scaled dipole moment $P/M^2$ vs scaled mass $M/\lambda$
($l=1$: dashed red);
(b) fundamental rotating solutions:
scaled angular momentum $J/\lambda^2$ vs scaled mass $M/\lambda$
(existence line: solid blue, critical line dashed blue).
Also shown are the extremal Kerr solutions (solid red).
}
        \label{Fig:domain}
\end{figure}

Before we turn to rotating black holes, let us briefly recall 
the properties of static black holes in this theory
\cite{Silva:2017uqg,Blazquez-Salcedo:2018jnn}.
The bifurcation points of the sets of static scalarized black holes
%obtained with quadratic coupling
from the branch of Schwarzschild black holes have been obtained
for the quadratic coupling function in \cite{Silva:2017uqg}, 
and they agree with those of the exponential coupling \cite{Doneva:2017bvd},
since the latter reduces to the quadratic coupling in the limit
of small scalar field.
The fundamental ($n=0$) static branch thus arises at the bifurcation point
$M/\lambda=0.587$, where the Schwarzschild solution developes a zero mode,
which turns into a first unstable radial mode for smaller $M/\lambda$.
At the $n$-th $(n>0$) bifurcation point, the Schwarzschild solution
developes the $(n+1)$-th unstable radial mode, 
while the associated $n$-th branch of scalarized black hole solutions 
possesses $n$ radial nodes %\cite{Silva:2017uqg}
and ($n+1$) unstable radial modes. 
Thus all scalarized black hole solutions are unstable, 
including the fundamental solution \cite{Blazquez-Salcedo:2018jnn}.
Furthermore we note that all branches of 
static scalarized black hole solutions
exist only in a small domain that decreases with increasing $n$,
as illustrated by the solid blue curves in Fig.~\ref{Fig:domain}(a),
%where their scaled scalar charge $Q/M$ is shown versus the scaled 
%coupling $\lambda/M$.
where their scaled scalar charge $Q/M$ is shown versus the scaled mass $M/\lambda$.
Note, that the figure also contains some angularly excitated solutions
(dashed red: $l=1$, green dots: $l=2$) discussed below.

In the following we present our results, 
starting with the domain of existence
of the fundamental rotating solutions. Subsequently, we address
rotating radial excitations and angular excitations.

\subsection{Fundamental rotating black holes}

The fundamental static black holes exist only in a small interval of
$M/\lambda$.  As these black holes are set into rotation, this small
interval shrinks further, as seen in Fig.~\ref{Fig:domain}(b) where
we plot the scaled angular momentum $J/\lambda^2$ as a function of the scaled
mass $M/\lambda$.  The boundaries of the domain of existence
correspond to the ``existence line'' to the left (solid blue line
starting at $M/\lambda=0.587$),
%
%\eb{Why plot $\lambda/M$ in the left panel? It's confusing. If we
%  stick with this choice, maybe we should quote both $M/\lambda=0.587$
%and its inverse in the text.}
%
and the critical (dashed blue) line on
the right where hairy solutions cease to exist: beyond the critical
line, there are only complex solutions for the scalar field.

The domain of existence resides fully within the domain of
existence of the Kerr black holes for the angular momenta studied.
As the scaled angular momentum $J/\lambda^2$ increases,
the domain of existence of the scalarized black holes
tends towards the set of extremal Kerr black holes.
We note that this behavior is rather similar to the case
of exponential coupling \cite{Cunha:2019dwb}.
However, in the latter case, the domain of static
scalarized black holes extends from the existence point
all the way to $M/\lambda=0$, yielding a large
domain of existence for small angular momenta $J/\lambda^2$.

\begin{figure}[t]
        \centering
\mbox{
\includegraphics[width=0.50\textwidth]{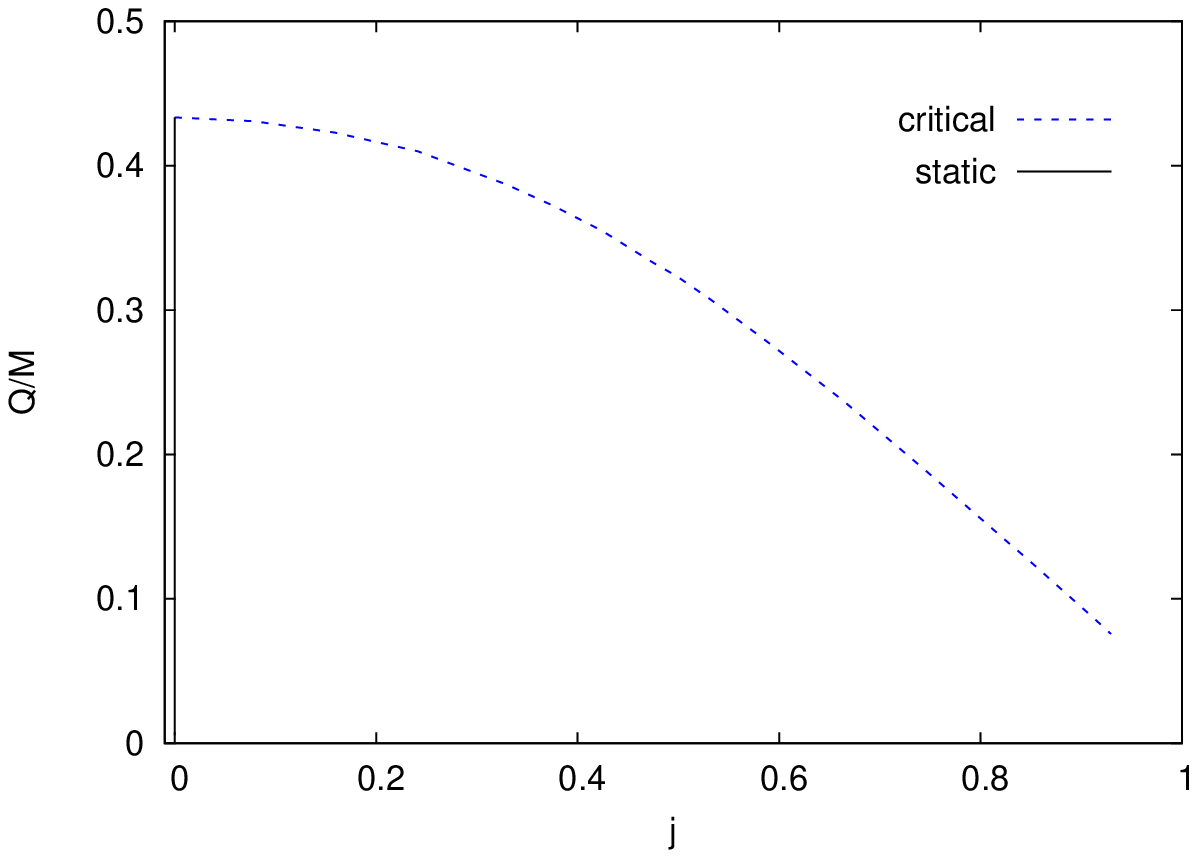}
\includegraphics[width=0.50\textwidth]{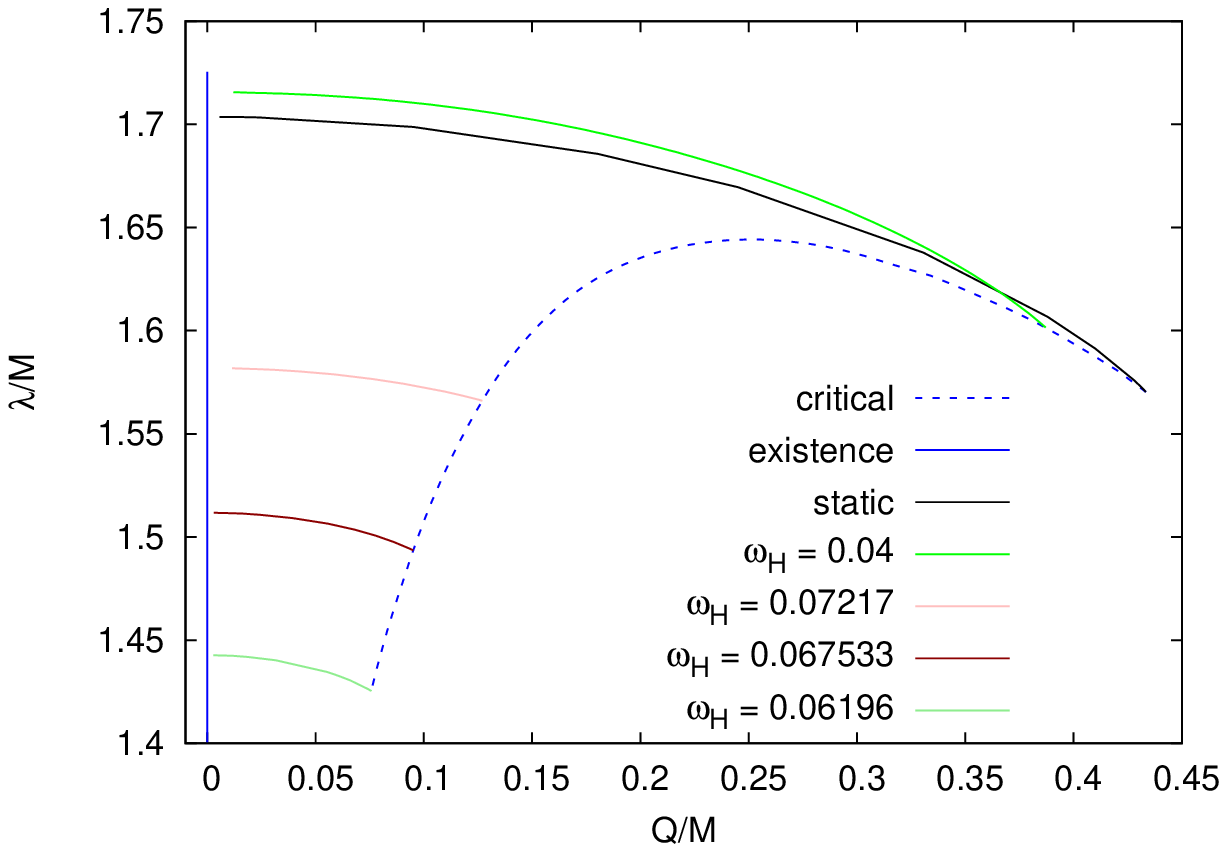}
}
        \caption{Domain of existence of 
fundamental rotating scalarized black holes:
(a) scaled scalar charge $Q/M$ vs scaled angular momentum $j=J/M^2$
(critical line: dashed blue, static solutions: solid black);
(b) scaled coupling constant $\lambda/M$ 
versus scaled scalar charge $Q/M$
(existence line: solid blue, critical line: dashed blue, static solutions:
solid black), for several fixed values of the horizon parameter
$\omega_{\rm H}=\Omega_{\rm H} r_{\rm H}$.}
        \label{Fig:domain2}
\end{figure}

In Fig.~\ref{Fig:domain2}(a) we illustrate the domain of existence 
by plotting the scaled scalar charge $Q/M$ as a function of the scaled angular momentum $j=J/M^2$. The scalar charge decreases
with the angular momentum, i.e. it decreases as we approach the 
extremal Kerr limit, where $Q/M=0$.
On the other hand, Figure ~\ref{Fig:domain2}(b) shows the domain of
existence with respect to the coupling constant. It also includes
curves of constant $\omega_{\rm H}$, which for constant $r_{\rm H}$
corresponds to a constant angular velocity of the horizon $\Omega_{\rm
  H}$ [cf. Eq.~(\ref{OmegaH})].
Interestingly, the static curve does not describe the boundary curve here,
but $\lambda/M$ can slightly exceed this curve for small 
rotation.
%
%\eb{Can we make the figures bigger? This one in particular was very hard to read on paper.}

\begin{figure}[t]
        \centering
\mbox{
\includegraphics[width=0.50\textwidth]{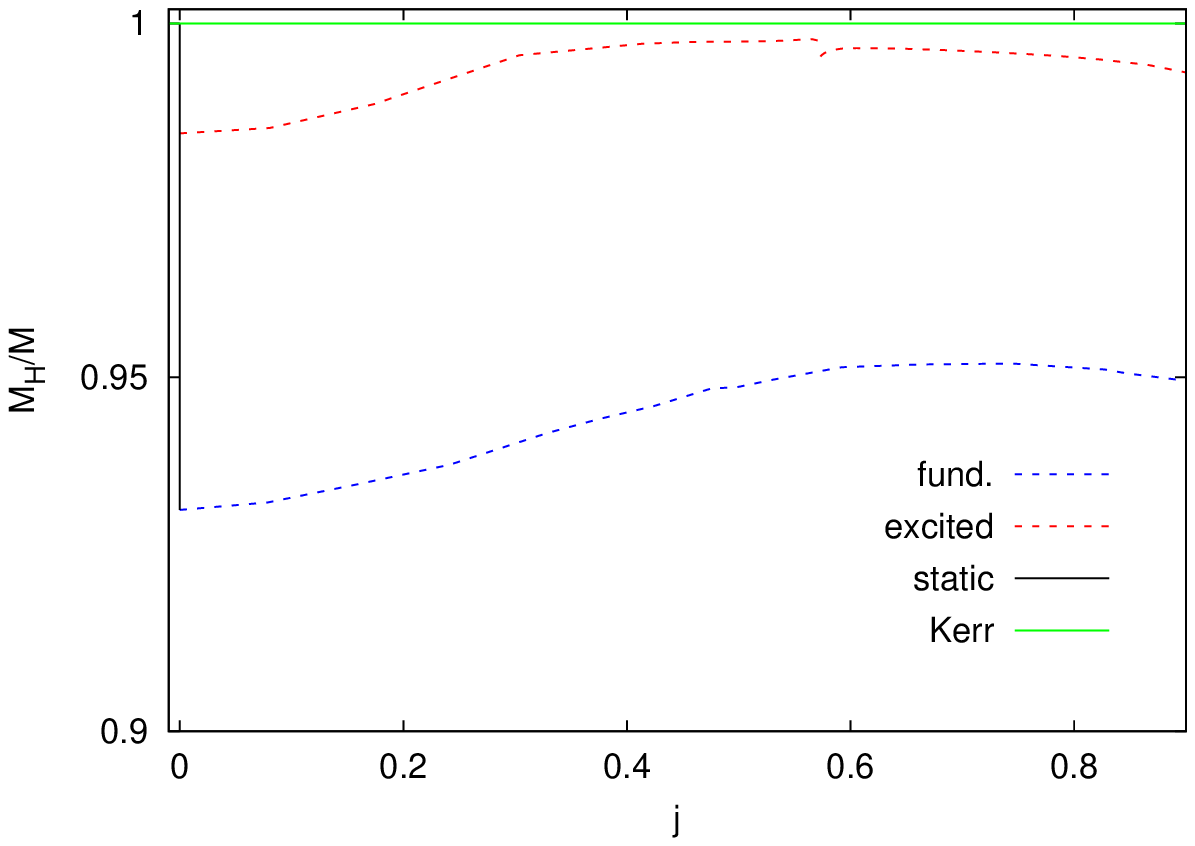}
\includegraphics[width=0.50\textwidth]{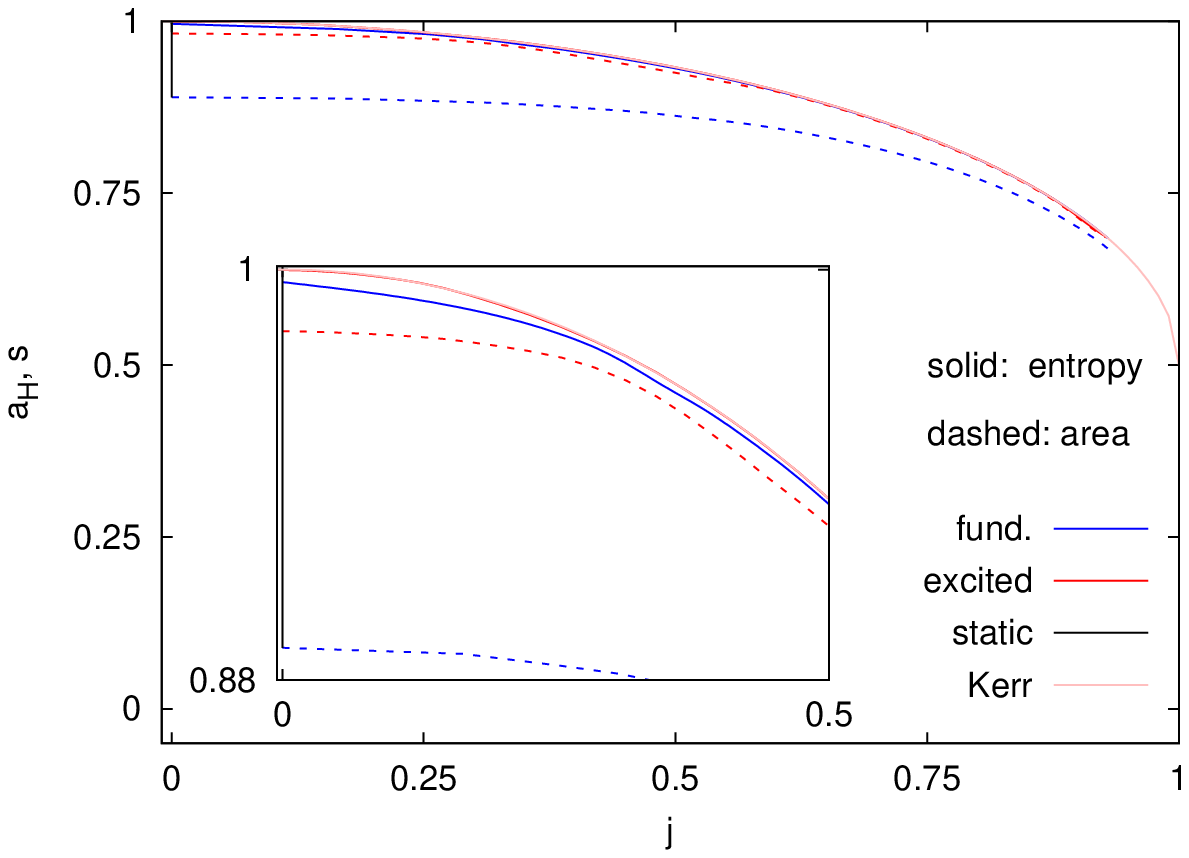}
}
        \caption{Domain of existence of
fundamental rotating scalarized black holes:
(a) scaled horizon mass $M_{\rm H}/M$ vs scaled angular momentum $j$
(critical line: dashed blue, static line: solid black, Kerr solutions: green);
%(b) scaled horizon area $a_{\rm H}=A/16 \pi M^2$ 
(b) scaled horizon area $a_{\rm H}=A_{\rm H}/16 \pi M^2$ 
and scaled entropy $s=S/4 \pi M^2$
vs scaled angular momentum $j$
(critical line: dashed blue, static solutions: solid black,
Kerr solutions: solid pink),
the inset illustrates the small difference in entropy.
The first radially excited solutions are also included
in the figures (red curves).}
        \label{Fig:domain3}
\end{figure}

As demonstrated in Fig.~\ref{Fig:domain3}(a),
the scaled horizon mass $M_{\rm H}/M$ is typically slightly smaller
than one.  
%with the relative difference decreasing with increasing angular momentum $j$, as the extremal Kerr solution is approached.
%
%
%\eb{Is this true? The critical (red) curve seems to have a local maximum around $j=0.7$, am I misreading it?}
%
Again, for large angular momenta this is similar to the exponential
coupling case, where, however, much
larger deviations are observed for small angular momenta \cite{Cunha:2019dwb}.
Figure \ref{Fig:domain3}(b) exhibits the 
%domain of existence for the scaled area $A_{\rm H}/M^2$.
domain of existence for the scaled area $A_{\rm H}/16 \pi M^2$.
%
%\eb{Uppercase ($A_{\rm H}$) or lowercase ($a_{\rm H}$)? Be consistent with the figure labels.}
%
The horizon area is smaller than in the Kerr case,
but nowhere are the deviations between the two very large.
As noted in \cite{Cunha:2019dwb}, large deviations are possible for Gaussian coupling functions,
and this can be used to put bounds on the coupling 
via the black hole shadow of M87 \cite{Akiyama:2019eap}.

Of considerable interest for such scalarized black holes
is the calculation of the entropy, which differs from the 
corresponding value obtained from the horizon area.
If the entropy for the scalarized solutions is larger than
the entropy for Kerr (or Schwarzschild) black holes, 
this is an indication that the scalarized solutions will be stable.
Here we do not expect stability, since static black 
holes are already known to be unstable.
Indeed, Fig.~\ref{Fig:domain3}(b) shows an explicit calculation of the
scaled entropy, which is smaller than in GR (as expected). However, the domain of existence for
the entropy of the scalarized solutions is extremely close 
to the case of Kerr black holes.
The difference is highlighted in the inset of Fig.~\ref{Fig:domain3}(b).
%
%\eb{I see no inset, are you redoing the figure?}

\subsection{Excited static black holes}

\begin{figure}[t]
        \centering
\mbox{
\includegraphics[width=0.50\textwidth]{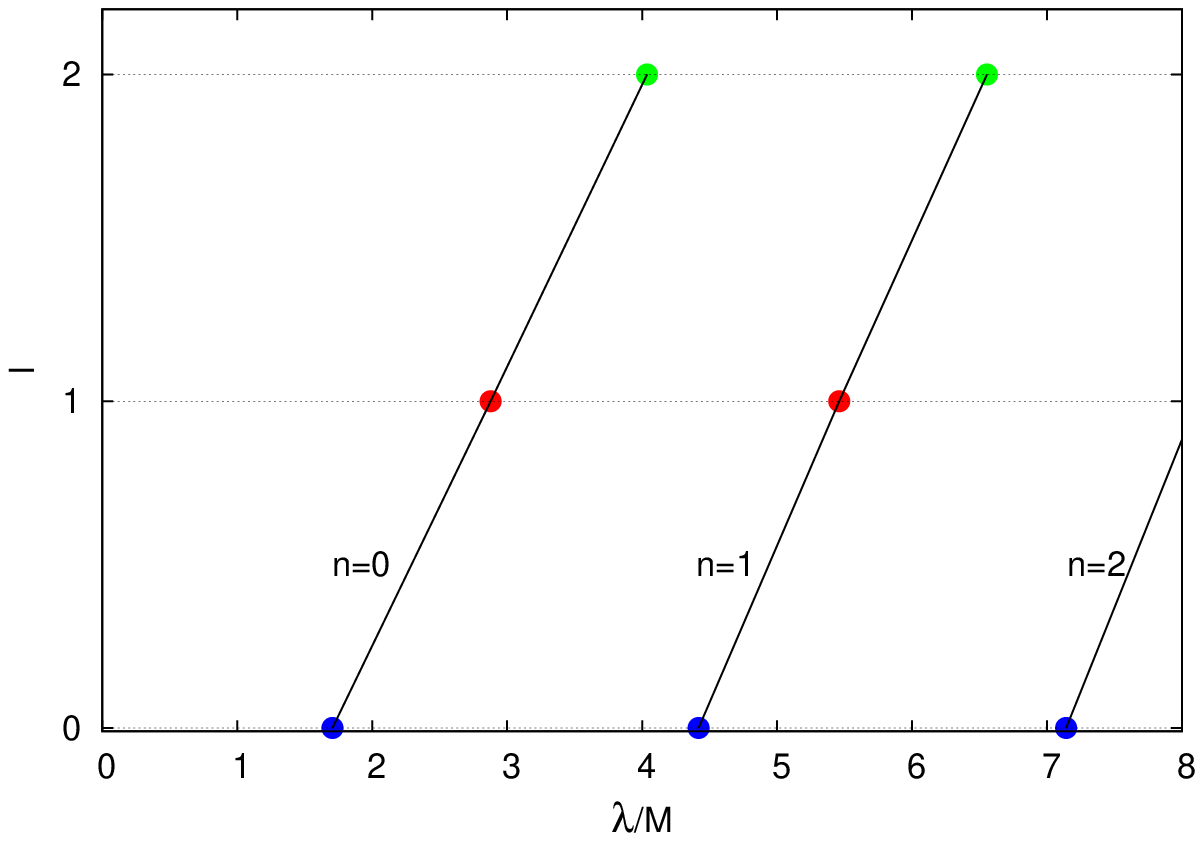}
}
        \caption{Bifurcation points of
excited static scalarized black holes:
scaled coupling constant $\lambda/M$ vs angular integer $l$
for the lowest radial excitations $n$;
$l=0$ (blue), $l=1$ (red) and $l=2$ (green).}

        \label{Fig:pattern}
\end{figure}

Scalarized black holes come in many variants.
Besides the well-studied radial excitations,
labeled by the integer $n$,
there are also angular excitations
and combinations of both.
In the static limit, this is easily seen when expanding the scalar field
in terms of spherical harmonics
\begin{equation}
\phi= \sum_{lm} f_l(r) Y_{lm}(\theta,\phi)
\end{equation}
involving the integers $l$ and $m$.
We exhibit the bifurcation points
of the lowest radially and angularly excited
static scalarized black holes in Fig.~\ref{Fig:pattern},
where the scaled coupling constant $\lambda/M$
is shown versus angular excitations, labelled by the integer $l$,
for the lowest radial excitations (labelled by the integer $n$).
Interestingly, the observed pattern of bifurcation points 
is highly regular, formed by adjacent rhomboids.

\begin{figure}[t]
        \centering
\mbox{
\includegraphics[width=0.50\textwidth]{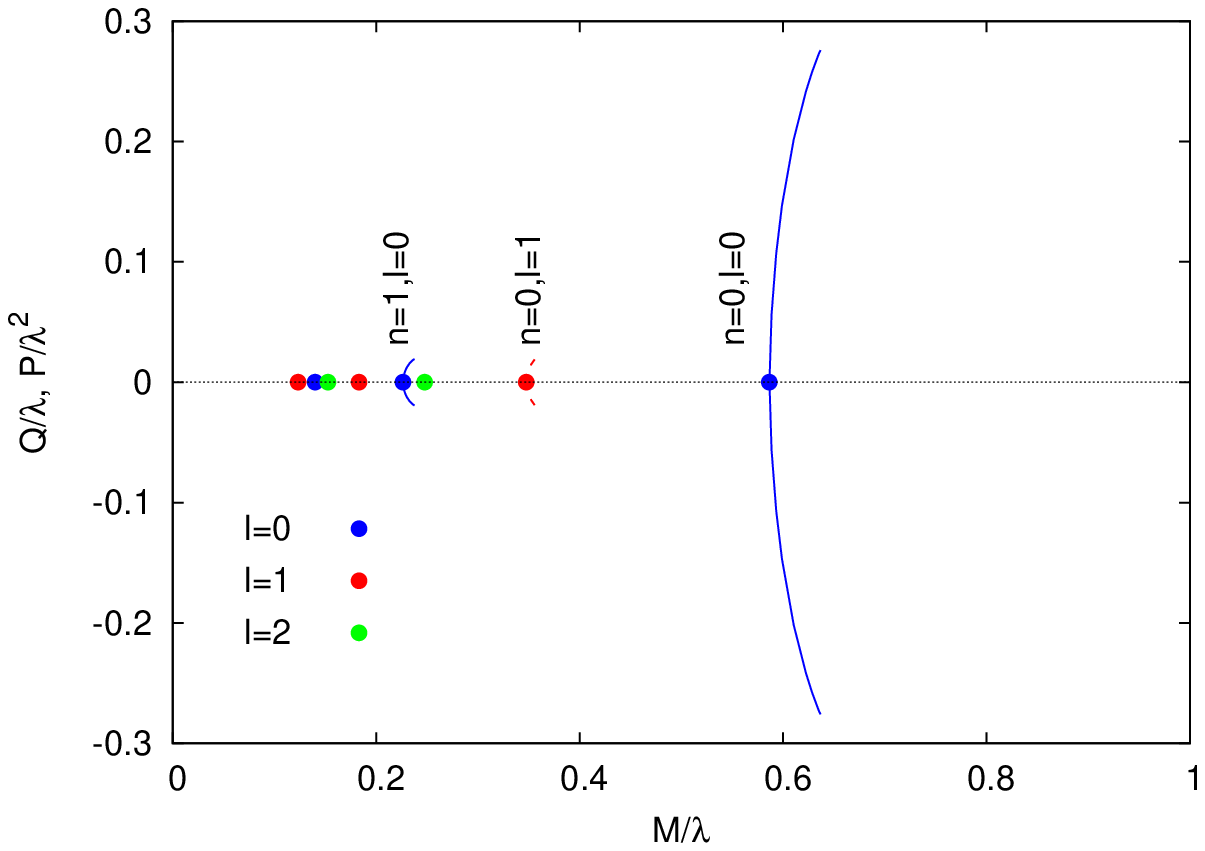}
\includegraphics[width=0.50\textwidth]{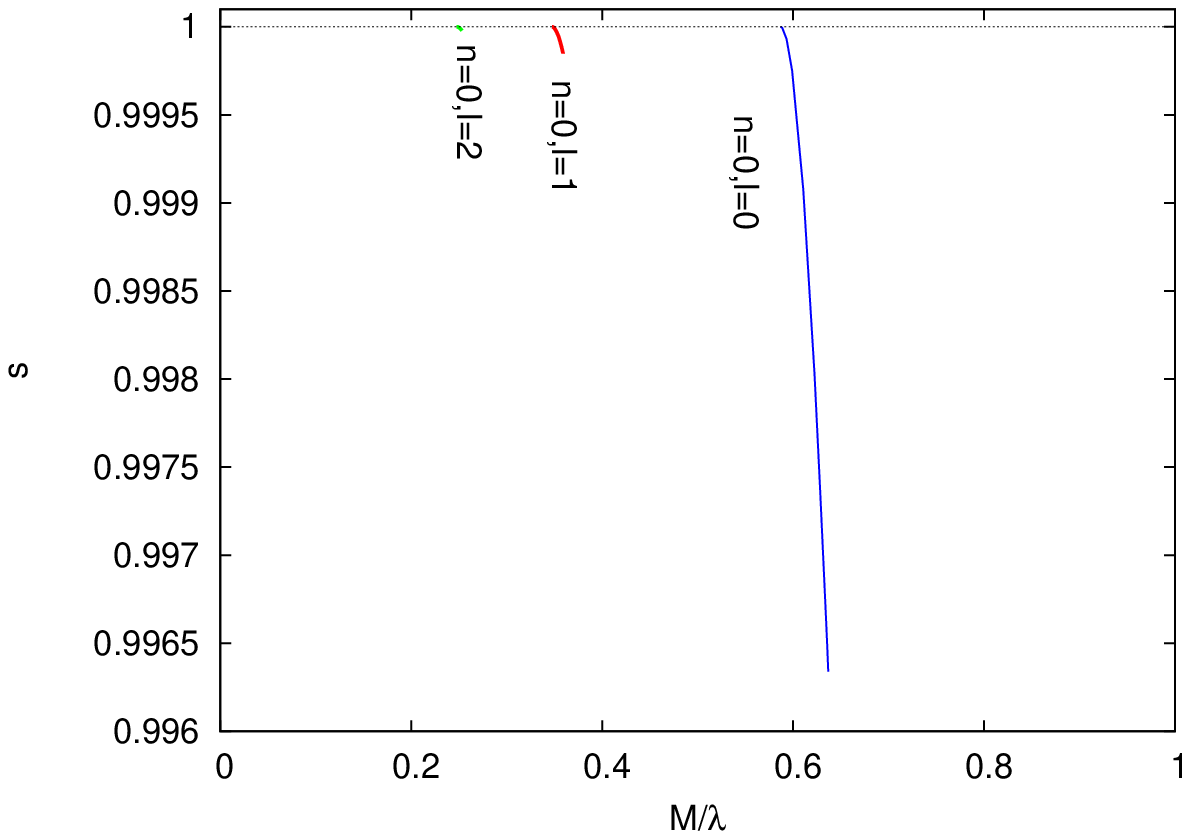}
}
        \caption{Excited static scalarized black holes:
(a) scaled scalar charge $Q/\lambda$ 
for even $l$ ($l=0$: solid blue, $l=2$: green),
and scaled dipole moment $P/\lambda^2$ ($l=1$: dashed red)
vs scaled mass $M/\lambda$;
(b) scaled entropy $s=S/4 \pi M^2$
vs scaled mass $M/\lambda$
for the lowest radial and angular exitations.
Bifurcation points: $l=0$ (blue), $l=1$ (red) and $l=2$ (green).
}
        \label{Fig:static2}
\end{figure}

Let us now consider the extension of the associated 
radially and angularly excited static solutions,
restricting attention to their existence lines, where
the scalar field equation is solved in the background
of the Schwarzschild black hole.
For these existence lines, Fig.~\ref{Fig:static2}(a) shows
the scaled scalar charge $Q/M$ (solid lines) and the 
%scaled dipole moment $P/M^2$ (dashed lines) as functions of the scaled coupling strength $\lambda/M$.
scaled dipole moment $P/M^2$ (dashed lines) as functions of the scaled mass $M/\lambda$.
In particular,
we show solutions with
$n=0,l=0$ (fundamental),
$n=0,l=1$ (one angular excitation)
$n=0,l=2$ (two angular excitations) and
$n=1,l=0$ (one radial excitation).
The dots in the figure indicate the bifurcation points 
from the Schwarzschild black holes for $l=0$ (blue), 
$l=1$ (red) and $l=2$ (green).
Note that solutions with $l=1$ possess a parity-odd scalar field,
therefore they do not carry scalar charge. 
The dipole term represents the lowest term In their asymptotic
expansion, so in the figure we plot their
dipole moment.
%
%\eb{Can we explain this better?}

We can ask whether static excited black holes are stable.
A glance at their entropy -- as shown in Fig.~\ref{Fig:static2}(b) --
indicates that not only the radially excited static solutions
are unstable \cite{Blazquez-Salcedo:2018jnn},
but also the angularly excited solutions should be unstable.

Of course, one might also consider static solutions, that in the
presence of backreaction will lose axial symmetry. 
Examples would be black holes labelled by the above 
set of 3 integers $(n,l,m)$, where the azimuthal integer $m\ne 0$, 
as in the solutions considered in \cite{Herdeiro:2018wub}.
For given integers $n$ and $l$,
these would possess a degenerate bifurcation point for all
allowed values of $m$, but give rise to a set of $l+1$ distinct 
families of scalarized black holes,
one with axial symmetry ($m=0$)
and $l$ without any continuous symmetry ($m\ne 1$).

\subsection{Excited rotating black holes}

\begin{figure}[t]
        \centering
\mbox{
\includegraphics[width=0.50\textwidth]{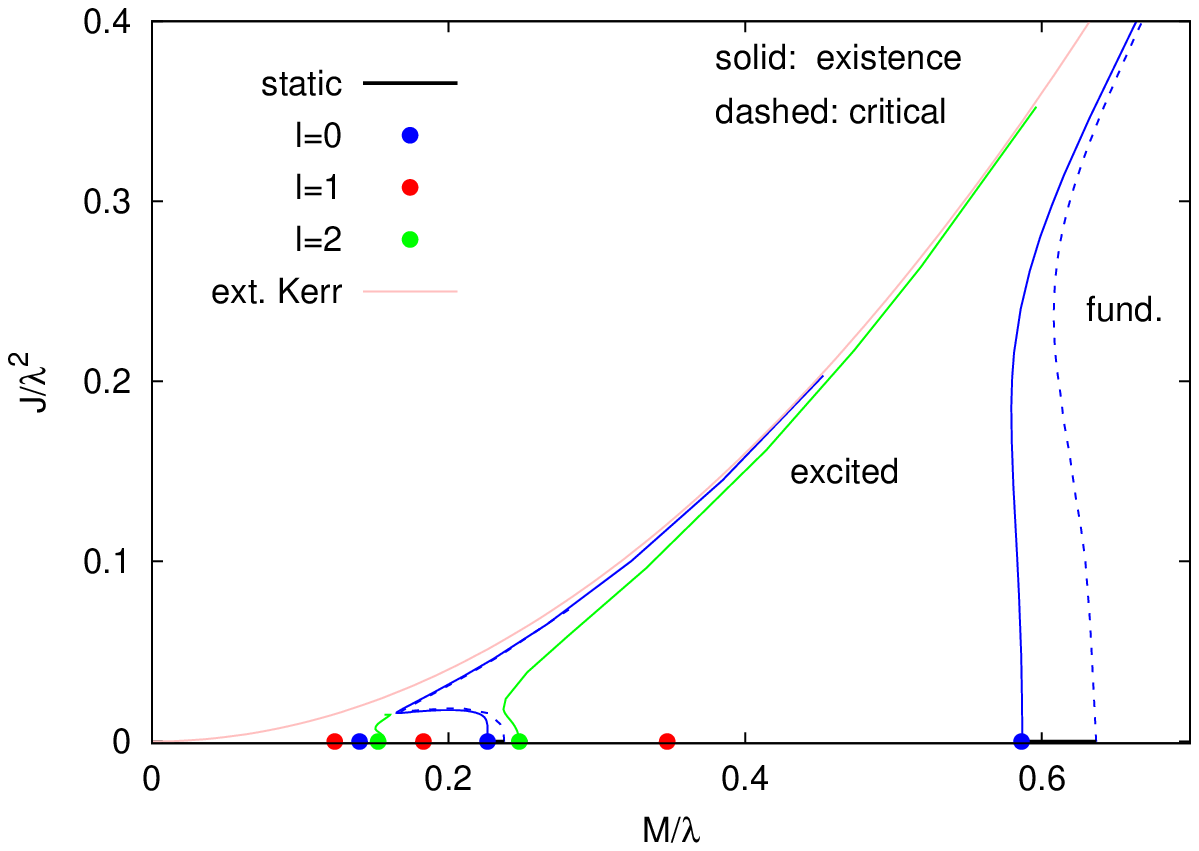}
\includegraphics[width=0.50\textwidth]{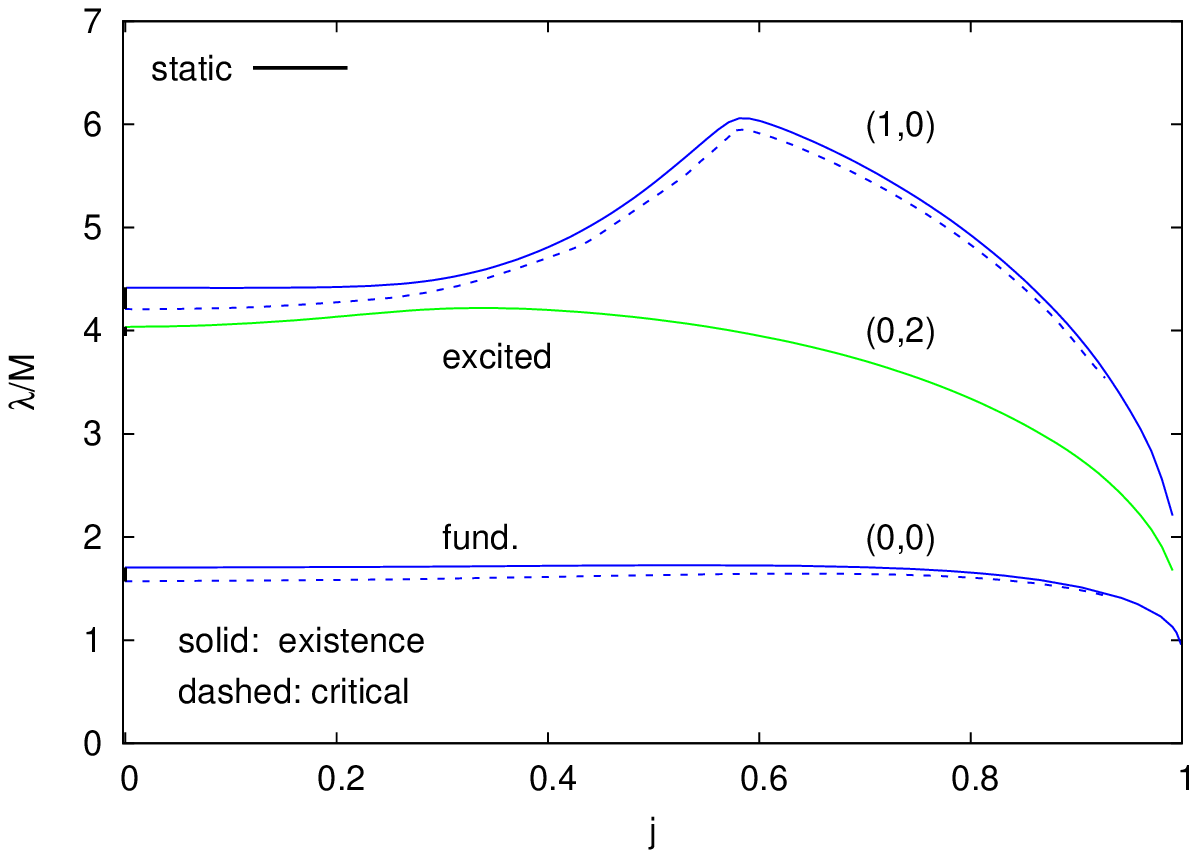}
}
        \caption{Excited rotating scalarized black holes:
(a) scaled angular momentum $J/\lambda^2$ vs scaled mass $M/\lambda$;
(b) scaled coupling constant $\lambda/M$ vs scaled angular momentum $j$.
Shown are $(n,l)=(0,0)$ (blue), $(n,l)=(1,0)$ (blue), $(n,l)=(0,2)$ (green)
(existence lines: solid, critical lines: dashed). 
}
        \label{Fig:domain4}
\end{figure}

Let us next consider excited rotating scalarized black holes.
We show the existence line for several such black holes
in Fig.~\ref{Fig:domain4}(a). In particular, 
in addition to the fundamental solution
for $l=0$, $n=0$ (blue),
the figure exhibits the existence line
for its first radial excitation  $l=0$, $n=1$ (blue),
together with the  existence line
of two branches of solutions starting from the bifurcation points
$l=2$, $n=0$ and $l=2$, $n=1$ (green).
As seen in the figure,
the existence lines $l=2$, $n=0$ and $l=0$, $n=1$ approach
the set of extremal Kerr solutions quite rapidly.
Also shown are the critical solutions with $l=0,n=0$ and $l=0,n=1$ (dashed).

In Fig.~\ref{Fig:domain4}(b) we plot the existence lines as functions of the
scaled coupling constant $\lambda/M$ and of
the scaled angular momentum $j$.
Besides the fundamental rotating solution $(n,l)=(0,0)$, we show 
also the first radially excited solution $(n,l)=(1,0)$,
and the second angularly excited solution $(n,l)=(0,2)$
\footnote{
The first radially excited branches of solutions
have also been included in Fig.~\ref{Fig:domain3},
where their scaled horizon mass, horizon area and entropy are shown.
In Fig.~\ref{Fig:domain3}(a) we notice a small gap,
that arises because in some small region of the parameters
$\left(\lambda, \omega_{\rm H} \right)$
the first radially excited solutions form multiple branches.
The study of these structures will be left to future work.}.
The short black lines on the axis indicate the
domain of existence of the static $(0,0)$ and $(1,0)$ solutions.
We have also obtained part of the existence lines
of further excited solutions like 
the second angularly excited solution $(n,l)=(0,2)$;
however, these remain challenging to fully map out.

\section{Conclusions}
\label{sec:conclusions}

Spontaneously scalarized black holes can exist when the coupling
function and coupling constant satisfy certain conditions.
In this paper we have studied curvature-induced scalarization mediated
by the presence of a GB term. In particular, we have focused on a
quadratic coupling function, extending previous studies of the fundamental ($n=0$)
nonrotating static solution and of its radial excitations.

We have mapped out the domain of existence of rotating generalizations
of these solutions, showing that the domain of existence is a small
band starting from the static solutions and extending to larger
angular momenta. This band shrinks as the angular momentum grows,
approaching the extremal Kerr solutions.  As for static black
holes, the entropy of these rotating scalarized black holes is smaller
than the entropy of Kerr black holes, suggesting that rotating
scalarized black holes should also be unstable.

We have further considered excited (static and rotating) solutions
considering also angular excitations with $l>0$. The bifurcation
points form a simple regular pattern in $(n,l)$, at least for small
values of $n$ and $l$.  Branches of excited black holes emerge from
these bifurcation points: this behavior is similar to the case of
static, charge-induced spontaneously scalarized black holes
\cite{Herdeiro:2018wub}.  Taking this similarity further, we
conjecture that there should also exist static EsGB black holes
without any continuous symmetry.

As long as axial symmetry is retained, the excited
solutions can also be set into rotation, forming stationary
sets of solutions. Here we have mainly explored the
existence lines of these radially and angularly excited
rotating black holes, but we also constructed solutions
with backreaction.

While we do not expect stable spontaneously scalarized black holes
for this theory, it should be possible to restore stability by adding
higher-order terms to the coupling function 
\cite{Doneva:2017bvd,Blazquez-Salcedo:2018jnn}
or by including a potential term $V(\phi)$  for the scalar field
\cite{Macedo:2019sem,Doneva:2019vuh}.
The latter approach is particularly attractive,
since these corrections would emerge naturally in an effective field
theory scenario.

\begin{acknowledgments}
The authors gratefully acknowledge support by the
DFG Research Training Group 1620  \textit{Models of Gravity}
and the COST Actions CA15117 \textit{CANTATA}
and CA16104 \textit{GWverse}. L.C. is thankful for the financial support obtained through the DFG Emmy Noether Research Group  under  grant  no.   DO  1771/1-1.
E.B. is supported by NSF Grant No. PHY-1912550, NSF Grant No. AST-1841358, NSF-XSEDE Grant No. PHY-090003, and NASA ATP Grant No. 17-ATP17-0225. 
This work has received funding from the European Union’s Horizon 2020 research and innovation programme under the Marie Skłodowska-Curie grant agreement No. 690904. This research project was conducted using computational resources at the Maryland Advanced Research Computing Center (MARCC). The authors are grateful to Eugen Radu for all the fruitful discussions.
\end{acknowledgments}

\end{document}